\documentclass[preprint,1p,authoryear,11pt]{elsarticle} 

\usepackage{graphicx}
\usepackage{amsmath}
\usepackage{amssymb}

\usepackage{multicol}

\usepackage{wasysym}

\usepackage[dvips,ps2pdf]{hyperref}

\hypersetup{
  pdfauthor    = Pasquale Tricarico <tricaric@psi.edu> and Mark Sykes <sykes@psi.edu>, 
  pdftitle     = The Dynamical Environment of Dawn at Vesta,
  pdfkeywords  = Dawn Vesta dynamics Tricarico Sykes ORSA celestial mechanics,
  pdfpagemode  = UseNone             , 
  pdfstartview = FitH                ,
  colorlinks   = true                , 
  linkcolor    = black               ,
  urlcolor     = black               ,
  pagecolor    = black               ,
  citecolor    = black               ,
  filecolor    = black               ,
}

\usepackage{soul} 

\usepackage[dvips]{color}

\newcommand{\memo}[1]{}

\newenvironment{packed_itemize}{
\begin{itemize}
  \setlength{\itemsep}{1pt}
  \setlength{\parskip}{0pt}
  \setlength{\parsep}{0pt}
}{\end{itemize}}

\journal{Planetary and Space Science}

\begin{document}

\begin{frontmatter}

\title{The Dynamical Environment of Dawn at Vesta}

\author[PSI]{P.~Tricarico\corref{cor1}}
\ead{tricaric@psi.edu}

\author[PSI]{M.~V.~Sykes}
\ead{sykes@psi.edu}

\address[PSI]{Planetary Science Institute, 1700 E.~Ft.~Lowell Rd.~Ste.~106, Tucson AZ 85719}

\cortext[cor1]{Corresponding author}

\begin{abstract}
Dawn is the first NASA mission to operate in the vicinity of the two
most massive asteroids in the main belt, Ceres and Vesta. This double-rendezvous
mission is enabled by the use of low-thrust solar electric propulsion. Dawn 
will arrive at Vesta in 2011 and will operate in its vicinity for approximately one year.
Vesta's mass and non-spherical shape, coupled with its rotational period, presents very interesting challenges 
to a spacecraft that depends principally upon low-thrust propulsion for trajectory-changing maneuvers. 
The details of Vesta's high-order gravitational terms 
will not be determined until after Dawn's arrival at Vesta,
but it is clear that their effect 
on Dawn operations creates the most complex operational environment for a NASA 
mission to date.
Gravitational perturbations give rise to oscillations in Dawn's orbital radius,
and it is found that trapping of the spacecraft is possible near the {\tt 1:1} 
resonance between Dawn's orbital period and Vesta's rotational period,
located approximately between 520 and 580 km orbital radius.
This resonant trapping can be escaped by thrusting at the appropriate orbital phase. 
Having passed through the {\tt 1:1} resonance, gravitational perturbations ultimately limit the minimum radius 
for low-altitude operations to about 400 km,
in order to safely prevent surface impact. 
The lowest practical orbit is desirable in order to maximize signal-to-noise
and spatial resolution of the Gamma-Ray and Neutron Detector and 
to provide the highest spatial
resolution observations by Dawn's Framing Camera and Visible InfraRed
mapping spectrometer. Dawn dynamical behavior is modeled in the 
context of a wide range of Vesta gravity models. Many of these models are
distinguishable during Dawn's High Altitude Mapping Orbit and the remainder are
resolved during Dawn's Low Altitude Mapping Orbit, providing insight into 
Vesta's interior structure. Ultimately, the dynamics of Dawn at Vesta identifies
issues to be explored in the planning of future EP missions operating in close
proximity to larger asteroids.
\end{abstract}


\end{frontmatter}


\section{Introduction}

The Dawn Discovery mission was successfully launched on September 27, 2007, and
is the first NASA science mission making use of solar electric propulsion (EP), 
enabled by the earlier Deep Space 1 technology demonstration mission \citep{1999BAAS...31.1127L}.
As a consequence of the efficiency of this low-thrust, low-acceleration system, 
Dawn is able to rendezvous with the two most massive objects in the asteroid 
belt, Vesta then Ceres. These targets were selected in order to study the 
earliest stages of planetary evolution for an object that formed dry (Vesta) 
and another that formed with substantial amounts of water (Ceres)
\citep{2004P&SS...52..465R}.

	Dawn's first target, Vesta, has a semi-major axis of 2.36 AU and is
located in the inner main asteroid belt. It is unique in having a basaltic crust
that has survived over the age of the solar system, providing important 
constraints on models of the collisional evolution of the asteroid belt
\citep[e.g.,][]{1985Icar...62...30D}. 
It has been spectroscopically linked to HED meteorites on the Earth 
\citep{1970Sci...168.1445M},
which represent approximately 6\% of all meteorite falls today 
\citep{1999nssy.book..351M}.
It has been inferred from those meteorites that Vesta is a 
differentiated object with an iron-rich core 
\citep{1984LPI....15..603N,1996LPI....27..407G,1996Icar..124..513R}.
Hubble observations of Vesta revealed an object 
with an equatorial radius around 289 km and polar radius of 229 km, but with 
a substantial impact crater covering much of its southern hemisphere and 
distorting its shape 
\citep{1997Icar..128...83Z,1997Icar..128...88T,1997Sci...277.1492T}.
This giant impact likely gave rise to the Vesta collisional family, which spans
the inner main belt from the $\nu_6$ secular resonance with Saturn on its inner edge to the {\tt 3:1} 
mean motion resonance with Jupiter, separating it from the outer main belt. 
Some Vesta material entering 
these resonances would have their orbits pumped into Mars- and eventually 
Earth-crossing orbits, resulting in the HED meteorites recovered on the Earth
\citep[e.g.,][]{1997M&PS...32..903M}. 
Dawn may provide connections between specific areas of Vesta's surface and HED (and possibly 
other) meteorites.

	Dawn executed a Mars gravity assist on February 17, 2009, to align its 
orbital inclination with that of Vesta. Dawn arrives at Vesta in July 2011 
when it will enter an initial orbit having a radius of 2700 km (Survey orbit), 
from which it will obtain a preliminary shape model using the Framing Camera 
(FC) and spectrally map the entire illuminated surface using the Visible 
InfraRed mapping spectrometer (VIR) 
\citep{2006AdSpR..38.2043R,2007EM&P..101...65R}.
Assuming the rotational pole of \cite{1997Icar..128...88T}, Vesta's obliquity is $27.2^\circ$.
Dawn arrives at Vesta at the time of maximum illumination 
of the southern hemisphere and its large crater. After completing the Survey 
orbit phase, Dawn uses its EP thrusters to descend to a High Altitude Mapping Orbit 
(HAMO) at approximately 950 km radius from which it will use the FC to map 
Vesta's surface and determine its global shape and local topography 
using stereophotoclinometric techniques \citep[e.g.,][]{2008M&PS...43.1049G}.
Dawn will then descend to its Low Altitude Mapping Orbit (LAMO) of 
around 460 km radius from which it will map Vesta's elemental composition 
using the Gamma-Ray Neutron Detector (GRaND). The Survey, HAMO, and LAMO 
phases are nominally 7, 27, and 90 days in duration \citep{2007AdSpR..40..193R},
but this is subject to further planning for the 8 month stay at Vesta, which 
may be extended to a year. Because Dawn uses its EP thrusters for orbit 
transfers, transitions between these different phases are expected to take around a 
month \citep{2007EM&P..101...65R}.

To maximize the science return from the mission, 
we are interested in determining
the lowest orbital radius from which Dawn can safely execute LAMO.
The closer to the surface we can make observations, the better the spatial 
resolution of FC and VIR observations. However, a lower LAMO most benefits
observations by GRaND. The spatial resolution of the GRaND instrument is 
approximately 1.5 times the altitude \citep{2003ITNS...50.1190P}. At nominal 460 km 
orbital radius for LAMO, this altitude will vary between 175 km near the 
equator and 231 km near the pole. By decreasing LAMO to 400 km, the number of 
GRaND resolved elements on Vesta increases by more than 50\%, 
improving our ability to identify geochemical units and relate them to HED
meteorites \citep{2010LPI....41.2299P}.
 Going lower also improves GRaND signal-to-noise and may enable an
accurate determination of Mg and Si, which are important discriminators among
the various rock types expected on Vesta 
(Prettyman, priv.~comm., and \citealt{2010LPI....41.2299P}).
However, lower orbits and corresponding decreased orbital periods
increases the need for desaturation of the spacecraft angular momentum,
increasing the operational burden, which is not explored here.

A Vesta gravity model is also greatly improved with reduced altitude, 
allowing for better detection of mascons and determination of Vesta's higher 
order gravitational terms. At HAMO, the gravity field is determined to at least
degree 4 and at LAMO this is expected to improve to at least degree 10 
\citep{2007EM&P..101...65R}. 
Depending on the accuracy of Doppler and Doppler-rate data,
simulations later in this manuscript show
that these numbers can be significantly improved,
reaching degree 10 at HAMO and degree 20 at LAMO.

While there are science benefits from a minimum radius LAMO, Vesta's large mass 
and deviation from sphericity raises the question of how its gravity field will 
constrain the lowest orbital radius at which Dawn can safely operate. 
In addition, the long transfer times between Survey, HAMO and LAMO mean that Dawn will be 
slowly transiting commensurabilities between its orbital period and Vesta's
rotational period, where perturbations on Dawn's orbit may be significant. 

With this work we explore the dynamics of the Dawn spacecraft in a polar orbit
within 1000~km from Vesta. The gravitational potential of Vesta is determined 
assuming diverse and extreme scenarios for its interior structure,
to ensure the dynamical environment is sufficiently explored and that the results
are representative of what the mission is likely to experience once there.
The orbital maneuver from HAMO to LAMO using EP to slowly spiral in is also simulated, 
to assess the effect of mean motion resonances.

\section{Modeling the Gravitational Field of Vesta\label{sec:modeling}}

In general, the gravitational potential of a body
with arbitrary shape and mass distribution
can be described using the spherical harmonics series \citep{1966tsga.book.....K,1995geph.conf....1Y}:
\begin{equation}
U(r,\theta,\phi) = \frac{G M}{r} \left[ 1 + \sum_{l=2}^{\infty} \sum_{m=0}^{l} \left(\frac{r_0}{r}\right)^{l} P_{lm}(\cos\theta) \left( C_{lm} \cos m\phi + S_{lm} \sin m\phi \right) \right] \label{eq:potential}
\end{equation}
where $G$ is the universal gravitational constant,
$M$ is the total mass of the body,
$\{r,\theta,\phi\}$ are the body-fixed barycentric spherical coordinates 
of the point where the potential $U$ is computed,
$r_0$ is an arbitrary reference radius usually corresponding to the radius of the 
Brillouin sphere, 
the smallest sphere enclosing the body,
$P_{lm}(\cos\theta)$ is the associate Legendre function.
Using this series, the Stokes coefficients $\{C_{lm},S_{lm}\}$ uniquely characterize the potential of the body.
In cases where the mass density distribution $\varrho(r,\theta,\phi)$ of a body is known,
then the coefficients $\{C_{lm},S_{lm}\}$ of the series in Eq.~(\ref{eq:potential}) 
can be determined by integrating 
over the volume $V$ of the body \citep{1995geph.conf....1Y}:
\begin{align}
C_{lm} &= \frac{(2-\delta_{m,0})}{M} \frac{(l-m)!}{(l+m)!} \int_V \varrho(r,\theta,\phi) \left( \frac{r}{r_0} \right)^{l} P_{lm}(\cos\theta) \cos m\phi \ \ \mathrm{d}V \label{eq:Clm} \\
S_{lm} &= \frac{(2-\delta_{m,0})}{M} \frac{(l-m)!}{(l+m)!} \int_V \varrho(r,\theta,\phi) \left( \frac{r}{r_0} \right)^{l} P_{lm}(\cos\theta) \sin m\phi \ \ \mathrm{d}V \label{eq:Slm}
\end{align}
The integration over the volume of the body is typically performed numerically,
i.e.~using Monte Carlo integration techniques \citep[i.e.,][]{1992nrca.book.....P}.
In the Monte Carlo integration, each integral $I = \int_V f \mathrm{d}V$ 
is approximated by $I \simeq V \langle f \rangle$,
where $\langle f \rangle$ is the mean value of the integrand inside the volume.
Both $V$ and $\langle f \rangle$ are computed by randomly sampling points inside the body's volume.
The nominal error on the result of a Monte Carlo integration is $\sigma_{I} = V \sigma_f/ \sqrt{N}$,
where $\sigma_f$ is the variance of the integrand,
and $N$ is the number of sample points used. 

The Stokes coefficients can be more readily compared and used in numerical
work when their magnitude is normalized \citep{1995geph.conf....1Y}:
\begin{equation}
\{\bar{C}_{lm},\bar{S}_{lm}\} = \sqrt{\frac{(l+m)!}{(2-\delta_{m,0})(2l+1)(l-m)!}} \{C_{lm},S_{lm}\}
\end{equation}
This normalization produced results in agreement the Stokes coefficients of the asteroid 433 Eros
produced by the NEAR mission, see \S\ref{sec:motion}.

In order to use this formalism to model Vesta's gravitational potential, 
and thus the dynamics of a spacecraft in its proximity,
we need a model for its shape and a model for its mass density distribution.

Analysis of Hubble Space Telescope (HST) observations of Vesta have yielded 
accurate determination of its size, shape and rotational state \citep{1997Icar..128...88T}. 
The overall shape of Vesta can be fit by a triaxial ellipsoid of radii 289, 280, 229, $\pm$5~km \citep{1997Icar..128...88T}.
The shape is not perfectly ellipsoidal, with departures of 15-20~km from the smooth ellipse,
and a large indentation with depth of 20-30~km and diameter of about 200~km in the southern hemisphere \citep{1997Icar..128...88T}.
The rotation period used for Vesta is of 5.3421288 hours \citep{1997Icar..128...88T},
and for the purpose of this work, we assumed Vesta to be a primary axis rotator.
The 3D shape model obtained by \cite{1997Icar..128...88T} was used as reference for the study presented in this manuscript.

Vesta is most likely a differentiated body. 
Mineralogical and isotopic data of HED meteorites suggests that 
heating, melting, formation of a metal core, a mantle, and a basaltic crust
took place in the first few million years 
of solar system history \citep[e.g.,][and references therein]{2002aste.conf..573K}.
Thermal modeling by \cite{1998Icar..134..187G} suggests that 
heating by $^{26}$Al would keep the mantle of Vesta hot for $\sim$~100 My.
It is possible that the mantle experienced a substantial if not complete melting
that resulted in the formation of a metal core \citep{2002aste.conf..573K}.
Results by \cite{1997Natur.388..854L} on the excess $^{182}$W 
measured on eucrites samples suggests that accretion,
differentiation, and core formation on Vesta took place in the first 5--15~My.
\cite{1997M&PS...32..825R} estimated the radius of the core 
using mass balance from the density of Vesta and a variable fraction of silicates,
with their best estimate of a core radius smaller than 130~km,
an olivine-rich mantle with thickness $\sim$~65--220~km,
and a crust with thickness $\sim$~40--85~km.
The placement and integrity of that core is in question given the large impact event
evidenced by the hemispheric southern crater. This impact may have fragmented
the core. It may have caused it to be effectively displaced from the center of figure. 
This motivates us to consider several scenarios for Vesta's interior mass distribution for
modeling its gravitational potential.

\begin{figure}
\begin{center}
\includegraphics*[width=\columnwidth]{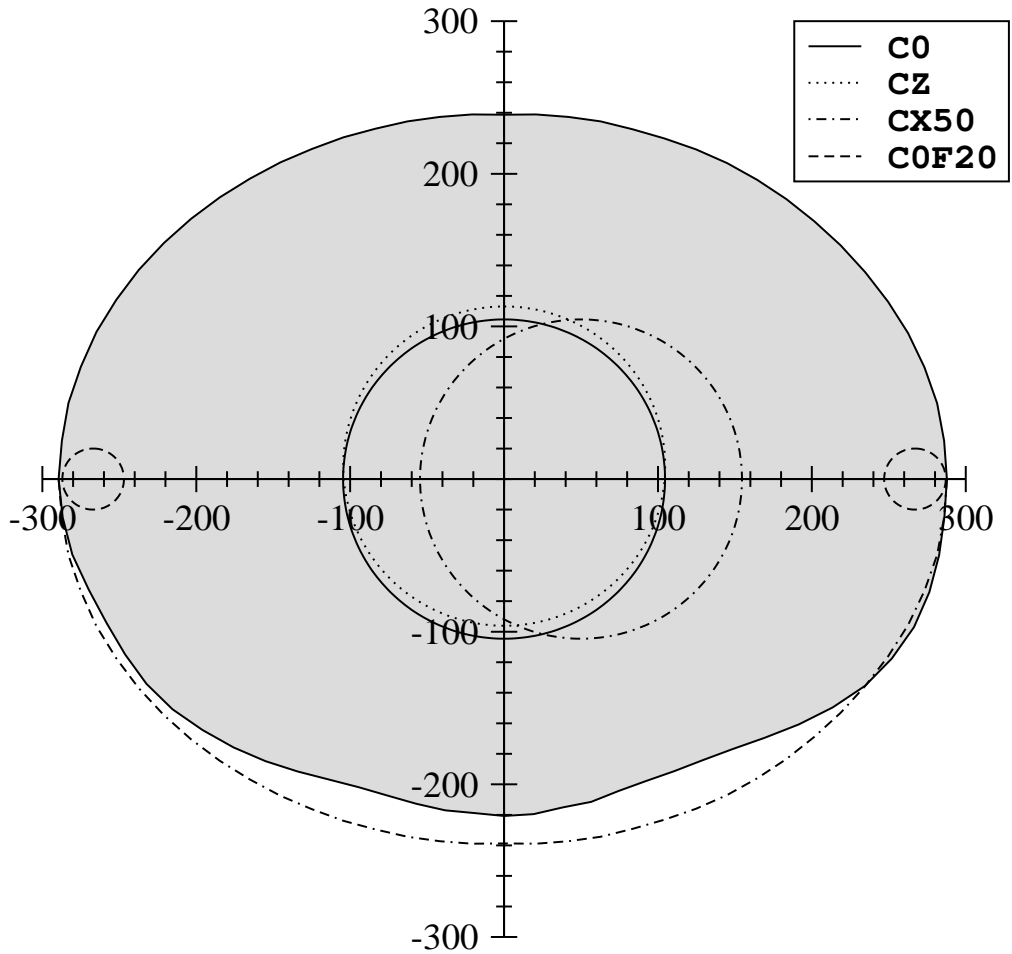}
\end{center}
\caption{Section of Vesta, shape model data by \cite{1997Icar..128...88T}. 
It fits within the envelope a triaxial ellipsoid with axes 577, 563, and 478 km (dashed-dashed-dotted line),
but deviates from that shape as a consequence of the nearly hemispheric impact crater covering the southern pole.
The characteristics of the four scenarios with a core are displayed.
The {\tt C0F20} scenario is characterized by a core at the origin, 
plus eight spherical fragments with 
radius 20~km equally spaced along the equator with the same density as the core.
North is up and scale units is in km.
}
\label{fig:shape}
\end{figure}

In this work we choose several scenarios for the mass density distribution of Vesta:
\begin{packed_itemize}
\item {\tt U} -- uniform mass density;
\item {\tt C0} -- a core centered on the origin of the tri-axial ellipsoid fit of the northern hemisphere;
\item {\tt CZ} -- a core centered in the same center of mass of the {\tt U} scenario;
\item {\tt CX50} -- a core offset by 50~km along the $x$ axis;
\item {\tt C0F20} -- a core centered on the origin, plus 8 fragments with the same density as the core, each of 20~km radius, equally spaced along the equator, just below the surface; 
\end{packed_itemize}
Our selection was dictated by the desire to explore a wide range of conditions,
from the very likely to the very extreme, in order to adequately characterize the dynamics of Dawn at Vesta.
The different internal mass distributions are
illustrated in Figure~\ref{fig:shape}.

Along with these scenarios, all using the known shape model,
we have included the {\tt EU} scenario, where the shape is that
of the smooth and symmetric triaxial ellipsoid, best fit of the northern hemisphere, with uniform mass density. 

In modeling the core or core fragments, we assume a density
of 7.90 g/cm$^3$, 
consistent with that of iron meteorites \citep{1998M&PS...33.1231C}.
Core size is constrained by conserving the total mass of Vesta and by fixing the 
density of mantle material to be 
3.12 g/cm$^3$, consistent with the grain densities of HED meteorites
(ibid.). Vesta is massive enough for its 
gravity to compress interior material to the point where it is one of only a
few asteroids thought to have no macro-porosity and little micro-porosity 
\citep{2002aste.conf..485B}.

\begin{figure}
\begin{center}
\includegraphics*[width=\columnwidth]{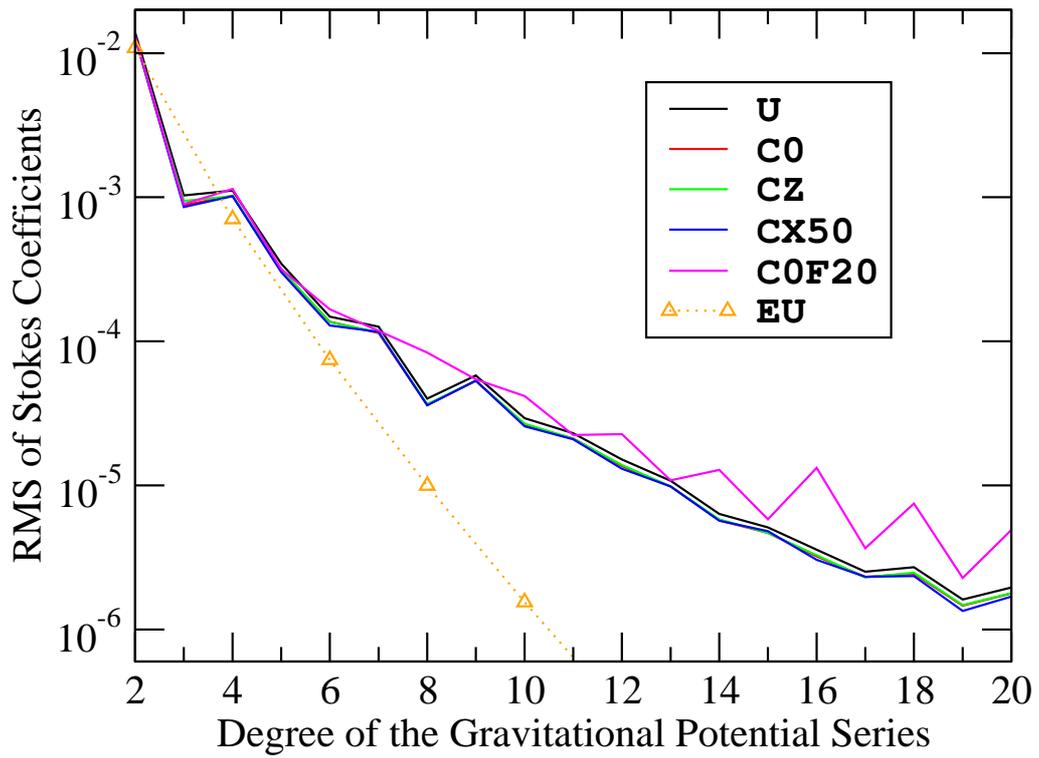}
\end{center}
\caption{RMS of the Stokes coefficients of the gravitational expansion as a function of the degree.
The {\tt EU} scenario is drastically different from the others (the power at odd degrees is zero),
and the {\tt C0F20} scenario presents larger power terms
at even degrees starting at degree 8, corresponding to a wavelength of $(\pi/4) R$ that is equal to the
spacing of the fragments on the equatorial plane.
}
\label{fig:power}
\end{figure}

For each scenario, we have derived an expansion of the potential,
using Eq.~(\ref{eq:Clm}) and (\ref{eq:Slm}) with $10^8$ sampling points,
sufficient to reach an integration error smaller than $10^{-6}$,
and the results are displayed in Table~\ref{table:scenarios}.
The root mean square (RMS) of all the normalized coefficients 
with a given degree $l$ is given by the formula
$[ \sum_{m=0}^l ( \bar{C}_{lm}^2 + \bar{S}_{lm}^2 ) / (2l+1) ]^{1/2}$ 
and provides a measure of the \emph{magnitude} 
of the degree $l$ of the gravitational expansion. 
If we interpret the expansion as a spectrum, then the characteristic wavelength of each degree 
is $\lambda_l = 2 \pi R / l$, with $R$ equatorial radius of the body.
In Figure~\ref{fig:power} we show the RMS of the coefficients for Vesta up to degree 20.
These coefficients were obtained using the shape of Vesta \citep{1997Icar..128...88T} 
under the scenarios considered.

Finally, Vesta's mass is derived from its influence on the 
motions of Mars and other asteroids, with a recent value of $1.35 \times 10^{-10}$~M$_\odot$
proposed by \cite{2009CeMDA.103..365P} and used throughout this work. 

\section{Simulating Dawn Motion Around Vesta\label{sec:motion}}

Given a gravitational potential and a rotational state for Vesta,
we simulate Dawn's orbital motion about Vesta under a wide range of initial conditions,
primarily sampling the initial radius and orbital phase.
We choose to run most of our simulations at the 8th degree when studying the
orbital motions of Dawn.
This choice is supported by our tests of the orbital evolution of Dawn at LAMO in all the scenarios considered,
where we have compared the dynamics of Dawn at degree 8 and 20. 
In our tests all the main features of the orbital evolution of Dawn were captured already at degree 8, and the difference 
in the orbital elements between the two cases were always of the order of 1\% over a period of 100 days at LAMO.

\begin{figure}
\begin{center}
\includegraphics*[width=\columnwidth]{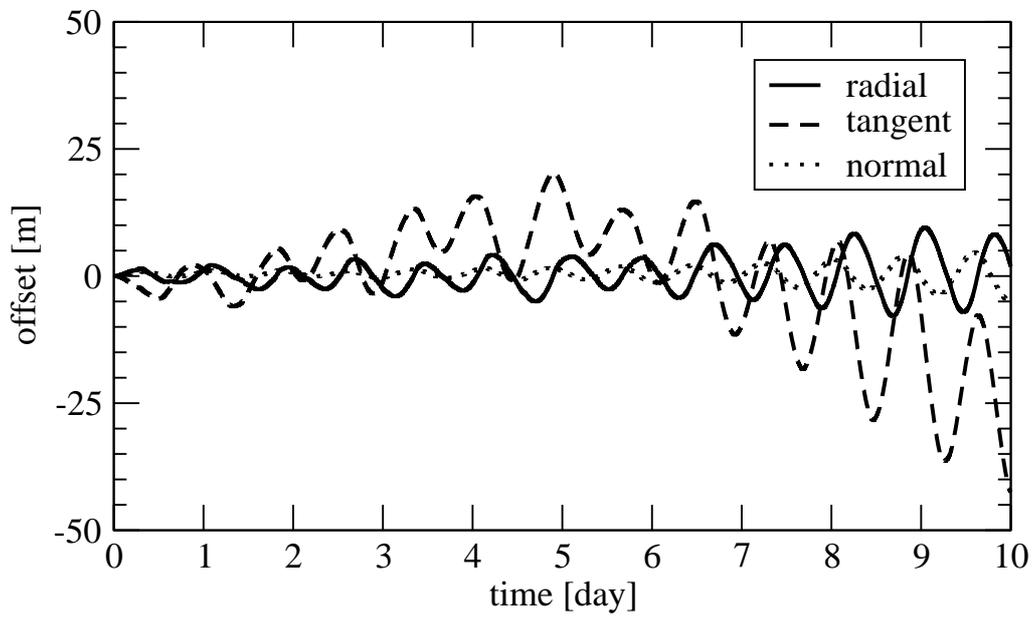}
\end{center}
\caption{Offset between the integrated NEAR trajectory using the ORSA code
and the nominal trajectory as retrieved from the SPICE kernels on PDS,
decomposed on its three spherical components: 
\emph{tangent} is parallel to the relative velocity vector,
\emph{normal} is orthogonal to the orbit plane.
}
\label{fig:NEAR}
\end{figure}

The numerical simulations are performed using a code specifically designed for this task.
This code is written in the C++ programming language, and makes very heavy use of a 
C++ framework designed specifically for celestial mechanics computations called ORSA\footnote{ORSA is open-source, available at http://orsa.sf.net.}.
The ORSA framework has been designed and continuously developed by Tricarico over 
the past decade, and has been used in several studies by Tricarico and collaborators.
Particularly relevant for this study is that the ORSA framework 
implements the interaction between extended bodies with arbitrary 
shape and mass distribution, using a formalism that  
is specifically designed to avoid transcendental functions,
that are computationally expensive \citep{2008CeMDA.100..319T}.
The numerical integrator used, 
based on the 15th order RADAU algorithm \citep{1985dcto.proc..185E},
is a variable step scheme that was originally developed for the integration 
of cometary dynamics, and that provides excellent accuracy 
for a wide range of dynamical problems.
This algorithm was modified to account for the effects of the primary's potential
on the spacecraft using the formalism in \cite{2008CeMDA.100..319T}.
This enhanced code was then validated for the dynamics in proximity of a rotating elongated primary,
using the reconstructed orbital data of the NEAR spacecraft
orbiting the asteroid 433~Eros, available in the Planetary Data System (PDS) 
\citep{2002Icar..155....3M,2002Icar..160..289K}.
As showed in Figure~\ref{fig:NEAR}, after 10 days the total offset 
between the reconstructed trajectory (PDS archived data) and trajectory integrated with our code
is still smaller than 50~m.
This also required the application of a simple model \citep{SRP} of radiation pressure 
acting on the NEAR solar panels, 
using a mass to projected area ratio of $34.5$~kg/m$^2$. 
The 10 day period was the longest period free of trajectory changing maneuvers at this close distance from the asteroid,
beginning on 7/14/2000 03:00 UTC, and corresponding to a 35~$\times$~37 km orbit.
As a comparison, the unmodeled stochastic accelerations present in the data,
with a magnitude of 7.5 m/day$^2$ and correlation time of 1 day \citep{2002Icar..155....3M},
would easily account for an offset in excess of 100 m over the same period,
so the numerical integration is well within the uncertainty of the reconstructed trajectory
of the spacecraft.

For Dawn, the effects of solar radiation pressure can be estimated using
the formalism in \cite{SRP}.
The projected area of Dawn consists of ten 1.6$\times$2.2 m$^2$ solar panels composing the array,
and a body of about 2$\times$2 m$^2$, for a total of 39.2 m$^2$.
The total mass of the spacecraft at Vesta will be of approximately $960$ kg 
(\cite{2006AcAau..58..605R} and Rayman, priv.~comm.), so the mass to
projected area ratio used here is $B \simeq 25$ kg/m$^2$.

\section{Assessing the Vesta Dynamical Environment}

The dynamics of a spacecraft orbiting within $\sim$~1000~km from Vesta
is constantly perturbed from a purely Keplerian orbit, in particular when crossing spin-orbit resonances.
When the orbital period of the spacecraft is close to a commensurability with the 
rotational period of Vesta, the effects of gravitational perturbations due to the high order terms 
of Vesta's gravitational potential can be amplified, 
and this leads to large perturbations in the semi-major axis and
eccentricity of the orbit of the spacecraft. 
This effect can be measured by monitoring the distance
between the spacecraft and Vesta (center-to-center) over a period (50 days) longer than
the time over which the perturbation acts (typically shorter than 10 days).
Over this monitored period, orbits in a spin-orbit resonance have a distance range (farthest minus closest)
significantly larger than orbits not affected by a spin-orbit resonance.

\begin{figure}
\begin{center}
\includegraphics*[width=\columnwidth]{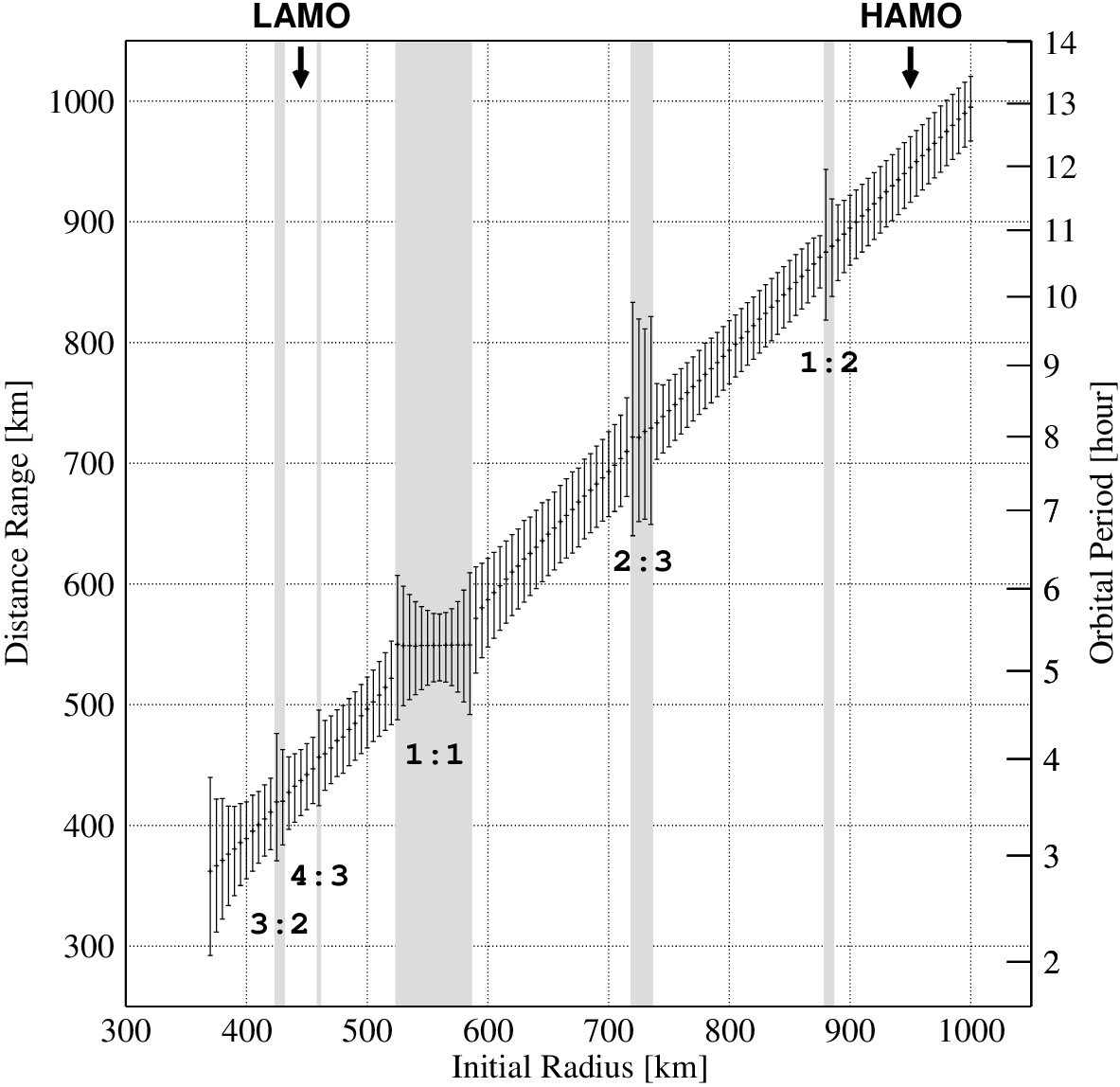}
\end{center}
\caption{Distance range as a function of the initial radius of a circular orbit, 
computed over a period of 50 days.
The central mark in each bar represents the median of the range.
The rotation period used for Vesta is of 5.3421288 hours \citep{1997Icar..128...88T}.
Five spin-orbit resonances have been identified and marked in the plot.
The {\tt 1:1} resonance affects the largest interval in initial radius,
but the strongest perturbations come from the {\tt 2:3} resonance.
The leftmost data point, with initial radius of 370~km, reaches a lowest distance 
just below 300~km, only a few km away from Vesta's surface.
The orbital radius of HAMO and LAMO is also marked.
}
\label{fig:range_R}
\end{figure}

We have sampled all the circular polar orbits 
with an initial radius between 370 and 1000~km at 5~km increments.
The dynamics of the spacecraft is numerically integrated over a period of 50 days,
subject to gravitational perturbations 
from the Sun, the planets, 
Vesta, and to solar radiation pressure.
We find that the two most important spin-orbit resonances are the {\tt 1:1} and the {\tt 2:3},
located at about 550~km and 720~km, respectively, see Figure~\ref{fig:range_R}.
The {\tt 1:1} resonance can affect orbits with an initial radius between about 500 and 590~km.
Similarly, the {\tt 2:3} resonance operates between 680 and 740~km.
The width of the resonances is not fully represented in 
Figure~\ref{fig:range_R} because only one initial condition per initial radius 
is sampled for the data in that figure.
Perturbations to the inclination of the polar orbits
were also observed, with oscillations of up to 4$^\circ$ total for resonant orbits,
while outside the resonances inclinations remained within one degree 
from polar for all distances larger than about 400~km.

Let us reiterate the fact that using an 8th degree gravitational potential is adequate for this study.
As discussed in \S\ref{sec:motion} the changes to the orbital elements at LAMO over 100 days 
due to degrees between 9 and 20 are
of the order of 1\%, so the dynamics of Dawn is completely captured in 8th degree simulations.
Also, the fastest resonances detected are the {\tt 3:2} and {\tt 4:3}, and the main resonant terms
for these resonances are respectively degree 3 and 4, so including all degrees up to 8 includes 
the first and second harmonic terms for the two fastest resonances.

\begin{figure}
\begin{center}
\includegraphics*[width=0.45\textwidth]{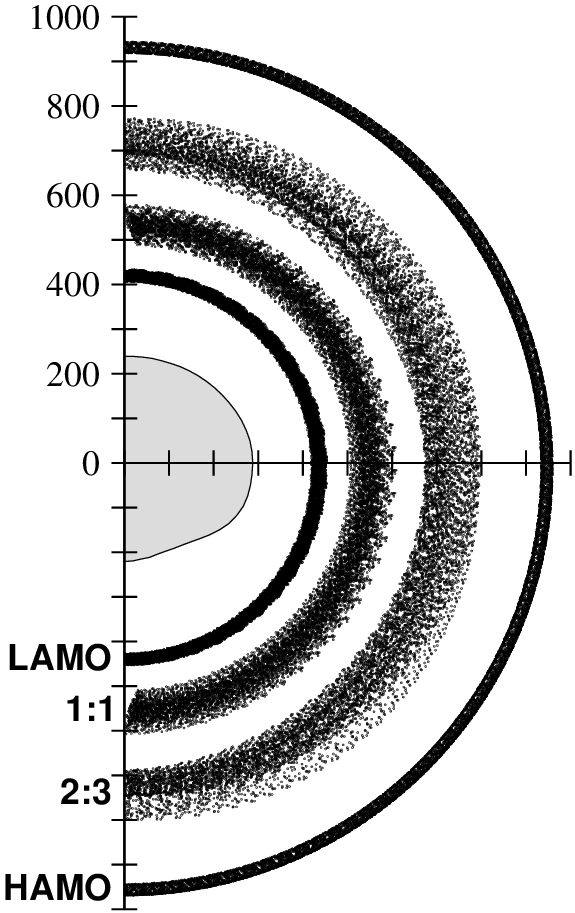}
\end{center}
\caption{Altitude variability for HAMO and LAMO configurations,
and for the spacecraft in {\tt 1:1} and {\tt 2:3} spin-orbit resonance.
Each cloud samples Dawn's trajectory over a period 50 days.
The north-south asymmetry in Vesta's shape forces an opposite asymmetry on Dawn's orbit,
which tends to get closer to the north pole than to the south.
North is up and scale units is in km.
}
\label{fig:polar}
\end{figure}

A graphical representation of the radial range variability is provided by Figure~\ref{fig:polar},
where we show in polar coordinates the \emph{cloud} of Dawn's dynamical evolution at HAMO, 
{\tt 2:3}, {\tt 1:1}, and LAMO over a period of 50 days. 
An interesting result is that the naturally perturbed trajectories of Dawn are not north-south 
symmetric, as they tend to get closer to Vesta at the north pole and farther away at the south pole.
When this effect is added to Vesta's already asymmetric shape,
it is clear that the altitude of the passages over the south pole will be even larger
than what the shape alone imposes to a spacecraft on a circular polar orbit.

\begin{figure}
\begin{center}
\includegraphics*[width=\columnwidth]{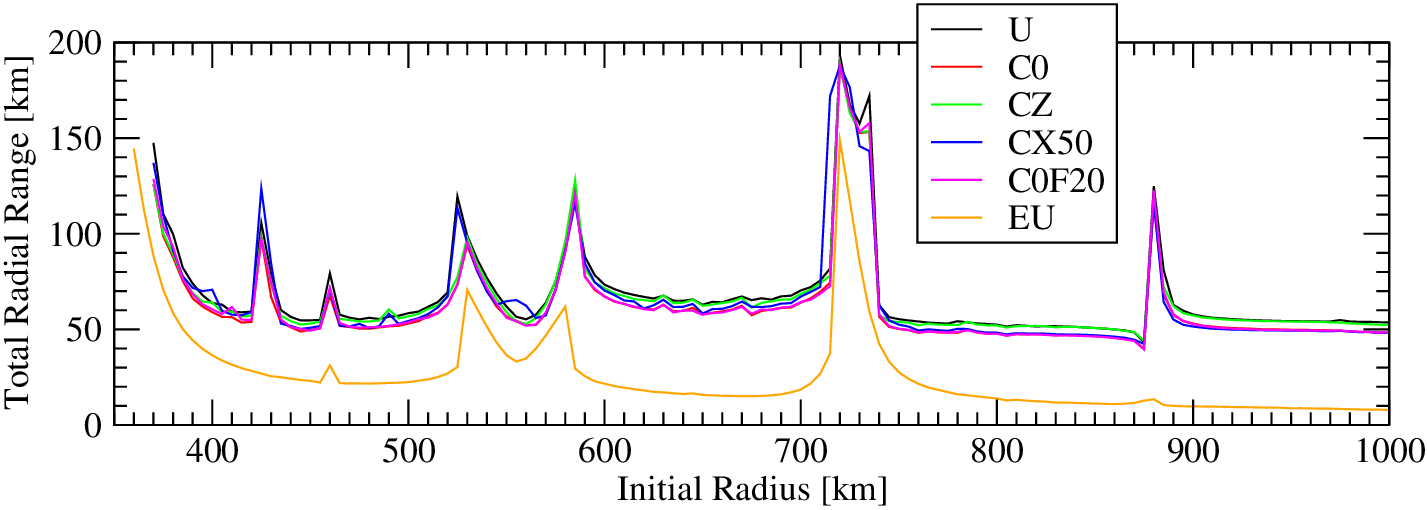}
\end{center}
\caption{Radial range for all the different scenarios considered.
With the exception of the {\tt EU} scenario, all scenarios
show a similar radial range independently of the internal mass distribution assumed.
Only one initial condition per initial radius was considered for each scenario.
For an analysis of the dependence on the initial conditions, see Figure~\ref{fig:twelve}.
}
\label{fig:allrange}
\end{figure}

We have also investigated to what extent the different interior structures 
considered in this manuscript affect the dynamics of Dawn,
using a similar approach of sampling circular polar orbits while monitoring radial range,
and the results are displayed in Figure~\ref{fig:allrange}.
We find it remarkable that over the range of reasonable diverse scenarios considered in this study,
the radial range pattern is mostly unaffected by the details of the interior structure.
An obvious exception is the {\tt EU} scenario, where the shape used is that of a smooth 
and symmetric triaxial ellipsoid and not the shape model based on \cite{1997Icar..128...88T} used in all the other scenarios.
This leads us to conclude that Dawn's dynamics, along with the multipole expansion of Vesta's potential,
depend primarily on Vesta's shape and only secondarily on the details of the interior structure
over the range considered, given the density constraints on mantle
and core material provided by HED meteorites and the measure of Vesta's mass. 
This is an advantage for the Dawn mission, because a detailed shape model will be constructed for Vesta
from Survey data which, in combination with gravity 
data obtained at HAMO, will help navigate the spacecraft through the {\tt 2:3} 
and {\tt 1:1} resonances when descending from HAMO to LAMO.

\begin{figure}
\begin{center}
\includegraphics*[width=\columnwidth]{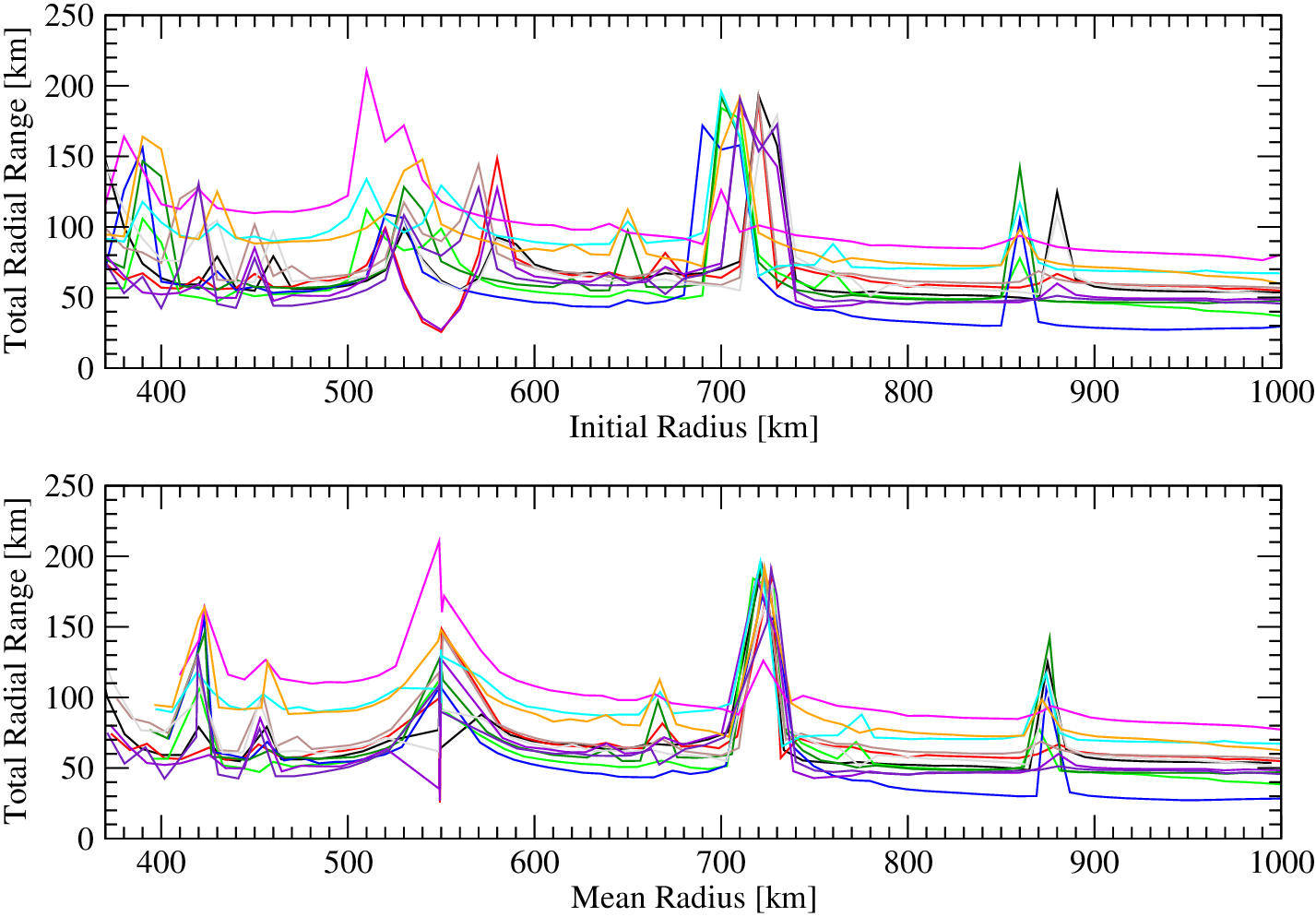}
\end{center}
\caption{Exploring the dependence to the initial conditions,
we have uniformly sampled circular orbits between 370 and 1000~km at $30^\circ$ intervals along the orbit.
Each line corresponds to a constant initial orbiting phase.
Results obtained using the {\tt U} scenario, and confirmed for specific 
cases using the other scenarios considered in the manuscript.
}
\label{fig:twelve}
\end{figure}

One more question is to what extent the details of the initial conditions,
such as initial orbital phase of Dawn and rotational phase of Vesta,
affect the dynamical evolution of Dawn.
We have performed once more the orbital sampling, but now for each initial radius we start 12
simulation at 30 degrees interval along the orbit,
using the {\tt U} scenario,
and the results are displayed in Figure~\ref{fig:twelve}.
We find that Dawn's dynamics depends strongly on the details 
of its initial orbital phase relative to Vesta's rotational phase.
In Figure~\ref{fig:twelve} we have plotted the same data first as a function of the initial radius,
and then as a function of the \emph{mean} radius, that here plays the role of an effective semi-major axis,
i.e.~the semi-major axis averaged over the simulated period of 50 days.
The former makes it easier to spot the initial value of the radius
at which a resonance can affect an orbit, 
while the latter clearly marks the exact position of the resonances.
The {\tt 1:1} and {\tt 2:3} spin-orbit resonances have very different effects.
At the {\tt 1:1} the largest magnitude of the radial range is reached 
for orbits starting at the edges of the resonance, 
with a minimum at its center located at about 550~km.
At the {\tt 2:3} the magnitude of the radial range is mostly a function of the mean radius:
orbits at the center of the resonance ($\sim$720~km) experience the largest perturbations,
reaching a radial range of 150-200~km and quickly decreasing moving away from it.
Out of spin-orbit resonances, the radial range still depends strongly on the initial conditions,
varying by up to a factor of 2.
While this sensitivity to the initial conditions was analyzed in detail using the {\tt U} scenario,
we have also studied specific cases using the other scenarios considered in the manuscript, 
confirming these results.

\begin{figure}
\begin{center}
\includegraphics*[width=\columnwidth]{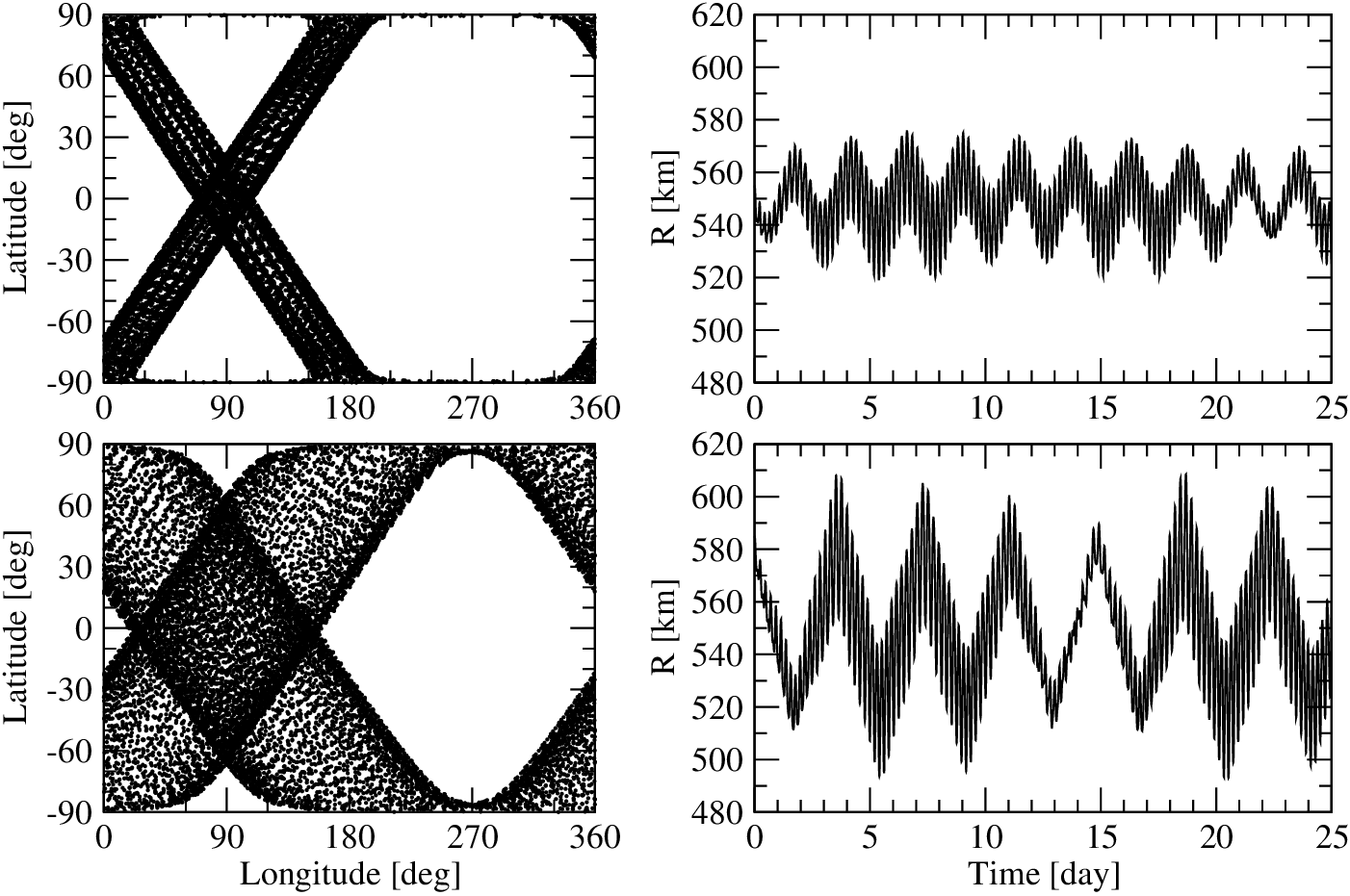}
\end{center}
\caption{Dynamics at the {\tt 1:1} resonance. 
Top, left: ground track of the spacecraft starting at 555~km, in the middle of the resonance, where the radial range is the minimum; 
right: orbit radius as a function of time for the 555~km simulation.
Bottom, left: ground track when starting at 585~km, at the external limit of the resonance, where the radial range is the largest; 
right: orbit radius for the 585~km simulation.
Zero longitude is at the longest equatorial radius of Vesta, so the {\tt 1:1} dynamics is characterized by a libration about the shortest equatorial radius of longitude $90^\circ$ or $270^\circ$.}
\label{fig:resonant}
\end{figure}

The effects of the {\tt 1:1} resonance are displayed in more detail in Figure~\ref{fig:resonant}. 
Depending on its initial radius, the spacecraft will librate with different amplitudes and periods about the shortest equatorial radius of Vesta. 

\begin{figure}
\begin{center}
\includegraphics*[width=\columnwidth]{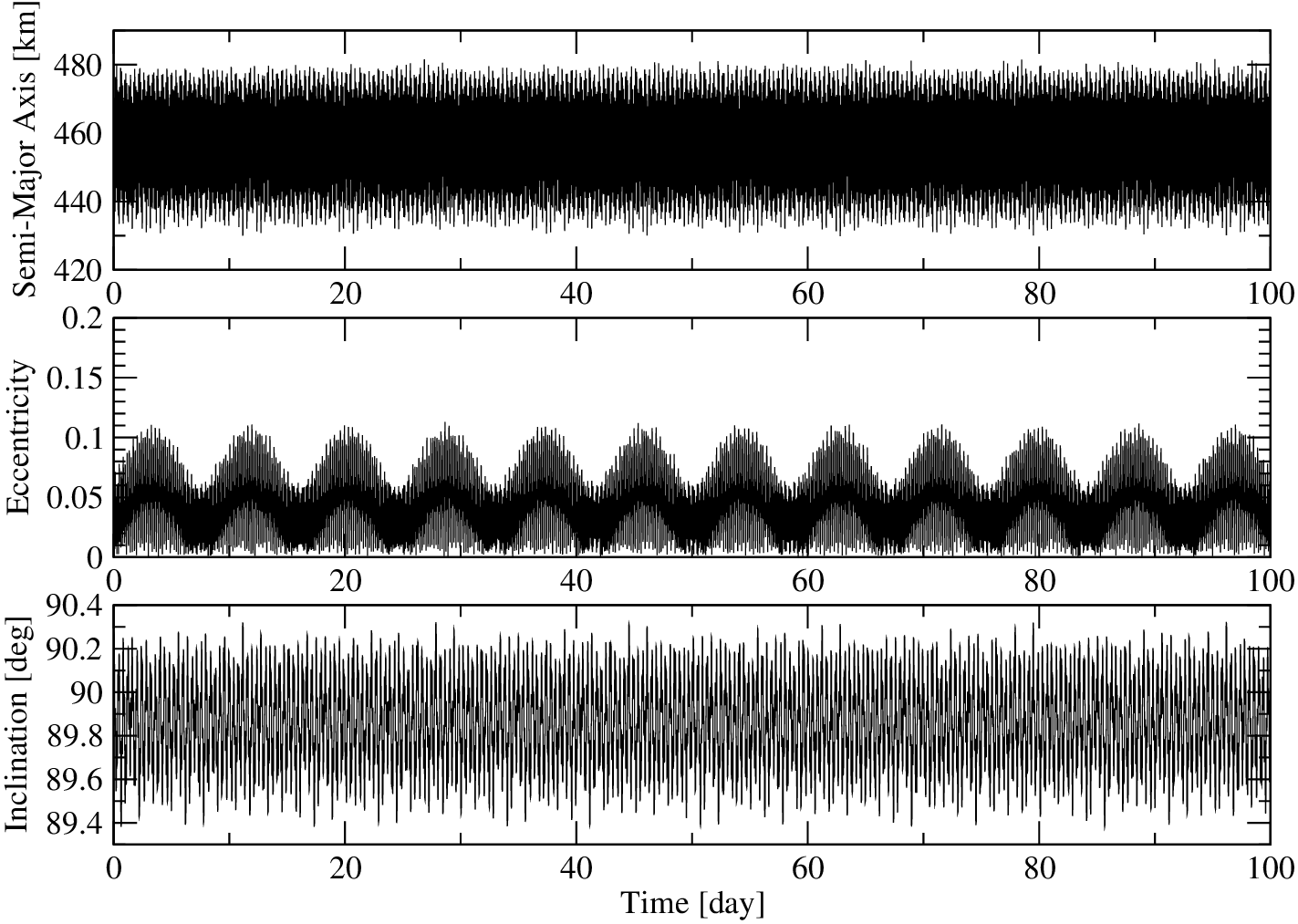}
\end{center}
\caption{Orbital evolution of Dawn at LAMO,
using the {\tt U} scenario and a gravitational potential including all terms up to degree 20, over a period of 100 days.}
\label{fig:orbital.LAMO}
\end{figure}

In Figure~\ref{fig:orbital.LAMO} we show the details of the orbital elements of Dawn at LAMO.
In this case we included all the terms up to degree 20 of the gravitational potential expansion,
to verify that no runaway effects are present even at this close distance,
over the expected period of performance of about 100 days.

\section{Operational Issues}

\begin{figure}
\begin{center}
\includegraphics*[width=\columnwidth]{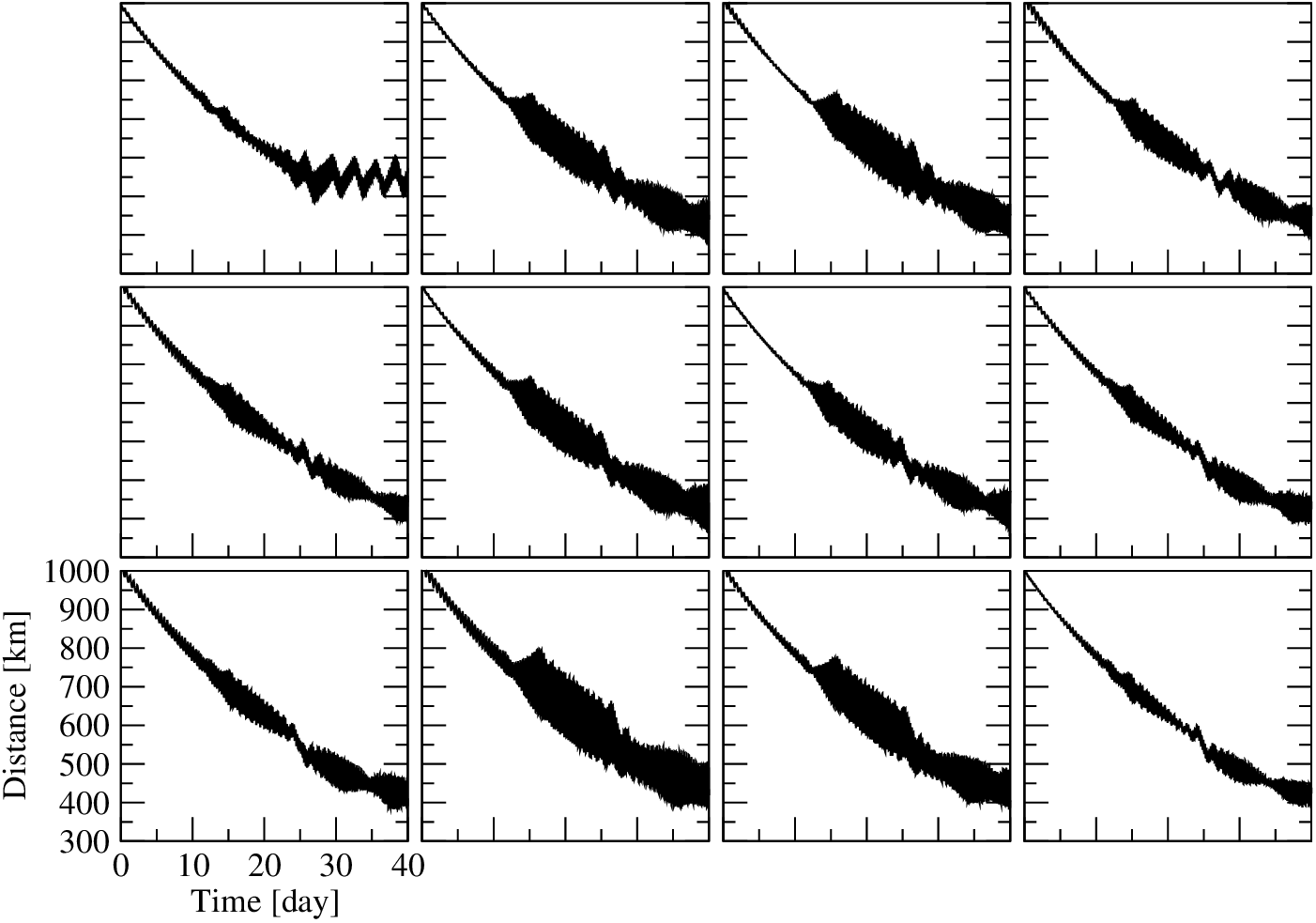}
\end{center}
\caption{Dynamics of the spacecraft spiraling in from a circular orbit of 1000~km radius.
The thrust is of 20~mN in all simulations, while the initial position is shifted by $30^\circ$ along the polar orbit.
Only in one simulation out of twelve the spacecraft gets trapped in the {\tt 1:1} spin-orbit resonance between 500 and 600~km.
All the other simulations reach the target 400~km orbit.}
\label{fig:grid}
\end{figure}

In order to better understand the effects of commensurabilities on Dawn's
operation, we simulated its slow descent from HAMO using 20 mN of thrust (at 
the low end of Dawn's thrust range, \citealt{rayman.2007}), which achieves 
transfer to nominal LAMO in about 30 days. An initial circular orbit
and continuous thrusting are assumed,
with the direction of thrust opposite to that of the velocity relative to Vesta.
The results of the simulations are shown
in Figure~\ref{fig:grid}. 

What is immediately obvious is the strong sensitivity 
to the starting position around the HAMO orbit for a
given Vesta sub-longitude. Different
initial Dawn orbital phases for a given Vesta rotational phase mean that the
azimuthal asymmetries in Vesta's gravitational field will be experienced at 
different phases and will have different effects. Away from commensurabilities the effects
are somewhat randomized in the aggregate. However, near the {\tt 1:1} commensurability
Dawn tends to pass over the same Vesta surface location twice for each orbit so the 
effects are reinforcing. 

The most interesting effect of Vesta's gravity field on Dawn is in the vicinity
of the {\tt 1:1} resonance. There are circumstances under which the spacecraft 
can be trapped in the resonance even while continuing to thrust. This is also 
very sensitive to the initial orbital/rotational phase. Out of twelve 
simulations, only one evidences trapping (Figure~\ref{fig:grid}) while most other 
phases exhibit an apparent temporary trapping. The trapping seems 
to occur when the Dawn spacecraft attempts to cross the {\tt 1:1} resonance 
near the longitude where Vesta's equatorial radius is shortest, 
as demonstrated when showing how the {\tt 1:1} resonance operates in
Figure~\ref{fig:resonant}. While trapped, Dawn's 
orbit librates about this point.
Trapping is not limited to scenarios using 20~mN thrust, 
as we have also observed that higher thrusting of 50~mN can exhibit a similar behavior.

\begin{figure}
\begin{center}
\includegraphics*[width=\columnwidth]{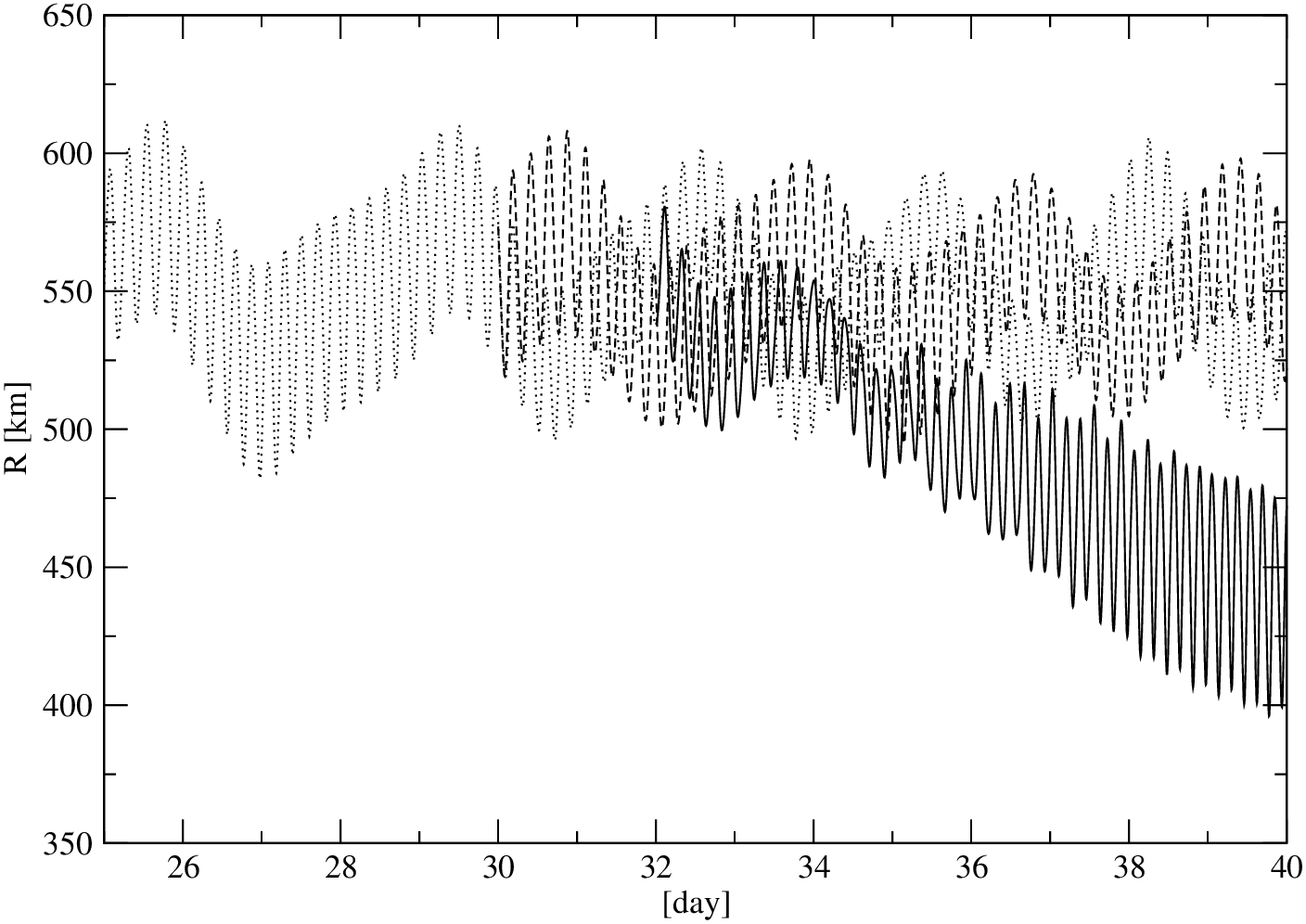}
\end{center}
\caption{Escaping the {\tt 1:1}. 
While Dawn is spiraling in thrusting at 20~mN it is captured into the {\tt 1:1} resonance (dotted line).
A first tentative to escape (dashed line) consists in increasing thrusting to 35~mN at time 30.0~days, but this does not succeed, as the spacecraft is still trapped, but with a different phase.
A second tentative (continuous line), thrusting to 35~mN at time 32.0~days and thus at a different phase,
successfully moves the spacecraft out of the resonance, towards LAMO.}
\label{fig:escape}
\end{figure}

Trapping can be escaped, but not by thrusting alone. Simulations in which the
thrust is increased from 20 to 35~mN show that the libration phase of such 
increased thrust is 
key (Figure~\ref{fig:escape}). In the meantime, the orbital motion within the 
resonance is well-behaved and does not appear to introduce a risk of spacecraft 
loss.

We also note how in more than half of the spiraling cases of Figure~\ref{fig:grid}
the amplitude of the radial oscillations increases significantly when crossing the {\tt 2:3}
resonance between 700 and 750~km,
pumping up the eccentricity of the orbit.
We anticipate that it might be necessary for Dawn to take this effect into account
and try to reduce the eccentricity of the orbit before reaching and crossing the {\tt 1:1} resonance
when migrating from HAMO to LAMO.

\section{Probing the Interior Structure of Vesta}

The interior structure of Vesta is manifested by the higher order terms of its
gravitational potential. The determination of these terms will depend upon the
reconstruction of the spacecraft position and orbital motion from coherent
X-band Doppler tracking \citep{2009DPS....41.5001A} and spacecraft-based observations of
Vesta against background stars.

\begin{figure}
\begin{center}
\includegraphics*[width=\columnwidth]{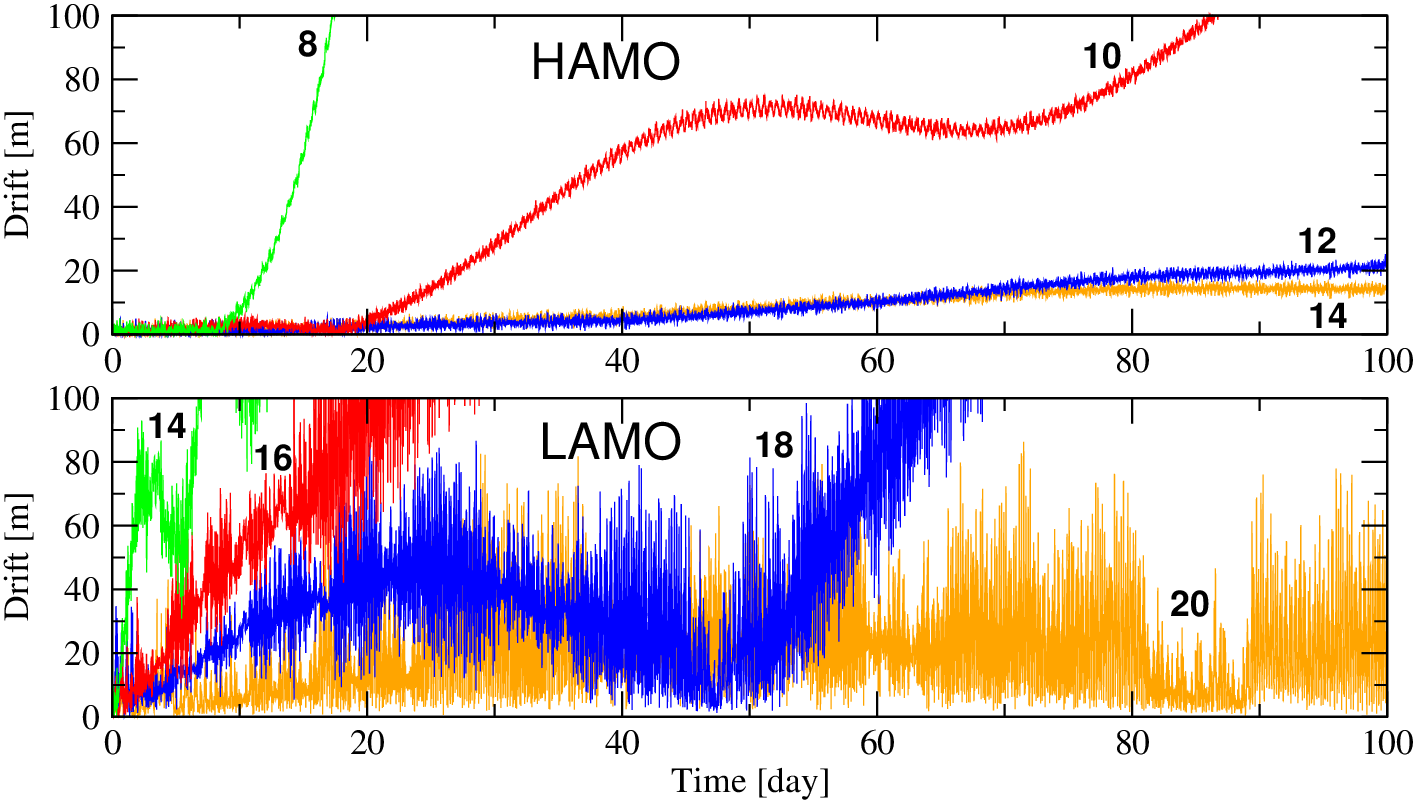}
\end{center}
\caption{Spacecraft drift due to the terms of a given degree 
in the expansion of the gravitational potential.
At HAMO (top) for degrees $\{8, 10, 12, 14\}$,
and at LAMO (bottom) for degrees $\{14,16,18, 20\}$.
The highest degree used in each curve is marked on the curve.
In order to effectively detect a drift between modeled and observed spacecraft dynamics,
the monitored period must not include any trajectory maneuver or thrusting.
}
\label{fig:degree}
\end{figure}

By simulating the dynamics of the spacecraft at different altitudes 
we can also determine the sensitivity to coefficients of a given degree
of the gravitational expansion at HAMO or LAMO,
as a function of the time free of trajectory maneuvers spent at each altitude.
As we show in Figure~\ref{fig:degree},
the perturbations due to terms of higher degree in the gravitational expansion
take a longer and longer time to produce measurable effects.
What we label as \emph{drift} in the figure is the offset between the predicted and observed 
position of the spacecraft, where the prediction was based on a model that included only terms up to the degree marked on each curve.
At HAMO, the effects due to the 8th degree become measurable after about 10 days,
those due to the 10th degree after about 20 days.
The effects of the 12th and 14th degree remain comparable over the 100 days period
simulated, and this reflects a condition of measurability in presence of strong correlation,
so the terms will not be determined uniquely at HAMO.
At LAMO, because of the reduced altitude, the sensitivity to higher degrees is evident,
and the effects due to the 14th and 16th degree are measurable within days,
while the effects of the 18th and 20th degree remain correlated over about 60 days.
This is in agreement with recent results by \cite{2009DPS....41.5001A},
who find that the 20th degree of the gravity field will be determined by Dawn.
Here we assume that the sensitivity of the deep space network is of the order of meters or tens of meters,
as is the case in similar conditions, when tracking spacecraft at Moon or Mars 
\citep[e.g.,][]{1998Sci...281.1476K,2001Icar..150....1K,2006Icar..182...23K}.

At this point it is an interesting exercise to determine the 
phase of the mission during which we can distinguish among 
all the scenarios considered.
The RMS of the difference between normalized coefficients of different scenarios 
listed in Table~\ref{table:scenarios} provides a measure of how much the expansion of any
two scenarios differs. We find that the difference between {\tt U} scenario
and any of the {\tt C} scenarios is of the order of $10^{-3}$ and is the largest observed.
The smallest difference observed is between {\tt C0} and {\tt CZ}, equal to $7\times 10^{-5}$.
Finally, the difference between all the remaining pairs of scenarios is of the order of $4\times 10^{-4}$.
Using Figure~\ref{fig:power} we can estimate the characteristic magnitude of the RMS of the coefficients 
(or their difference) of a given degree, and then using Figure~\ref{fig:degree}
we can assign a timescale to each degree. 
What we obtain is that RMS coefficients of magnitude $10^{-3}$
correspond to the magnitude of terms of degree 4;
coefficients of magnitude $4\times 10^{-4}$ correspond to the magnitude of terms of degree 5 to 6;
coefficients of magnitude $7\times 10^{-5}$ correspond to the magnitude of terms of degree 8 to 10.
Degrees below the 8th are not displayed in Figure~\ref{fig:degree}, but it is clear that 
data necessary to reach sensitivity below that necessary to determine the gravitational
terms of degree 8 will be acquired within the first few days of HAMO.
This is the time over which will be possible to distinguish between all the scenarios
except {\tt C0} and {\tt CZ}. To distinguish between the latter two, 
data for up to 20 days at HAMO will be necessary. 
So going back to the original question we can say that if Vesta's interior
was described exactly by one of the scenarios considered, Dawn would be able to
indicate which one this is within the first 20 days of HAMO.

\section{Conclusions}

Vesta presents novel and interesting operational challenges to the Dawn mission. 
Because of the low-thrust electric propulsion system, Dawn will pass
through Vesta spin/Dawn orbit commensurabilities very slowly as it moves from its High
Altitude Mapping Orbit to Low Altitude Mapping Orbit,
maximizing the effects of the perturbation. 
In addition to the rapid oscillations in Dawn's orbital radius as a consequence of
Vesta's complex gravity field, there is the potential that Dawn could be 
trapped near the {\tt 1:1} resonance, as it slowly decreases its radius through
550 km. Dawn can escape trapping by increasing thrust at the appropriate orbit 
libration phase. This is not an issue that is expected to recur at Ceres, which
is observed to have a simple oblate spheroid shape. 
Once through this resonance, Dawn can continue to decrease its orbital radius to around 
400 km (60 km less than the current LAMO) before the effect of perturbations
begin to progressively increase orbital radial oscillations until impact with the
surface becomes a hazard as Dawn reaches an average orbital radius of around
370 km.

\section*{Acknowledgements}

This work is supported by subcontract from UCLA to the Planetary
Science Institute under NASA prime NNM05AA86C for the NASA Dawn Discovery Mission.
We thank Tom Prettyman for discussions on GRaND performance and capabilities.
We also thank Stacy Weinstein-Weiss for pointing out practical 
issues such as de-sats in Dawn operations.
This is PSI contribution \#483.


{
\begin{table*}
\begin{center}
\footnotesize
\begin{tabular}{c|rrrrr|r|l}
           & {\tt       U} & {\tt    C0} & {\tt    CZ} & {\tt  CX50} & {\tt C0F20} & {\tt    EU} \\
\hline
$\rho_{m}$ &   $    3.411$ & $    3.120$ & $    3.120$ & $    3.120$ & $    3.120$ & $    3.411$ & g/cm$^3$ \\
$\rho_{c}$ &           --- & $    7.900$ & $    7.900$ & $    7.900$ & $    7.900$ &         --- & g/cm$^3$ \\
$R_{c}$    &           --- & $  104.589$ & $  104.589$ & $  104.589$ & $  102.601$ &         --- & km       \\
$x_{c}$    &           --- & $   +0.000$ & $   +0.107$ & $  +50.000$ & $   +0.000$ &         --- & km       \\
$y_{c}$    &           --- & $   +0.000$ & $   -1.145$ & $   +0.000$ & $   +0.000$ &         --- & km       \\
$z_{c}$    &           --- & $   +0.000$ & $   +8.503$ & $   +0.000$ & $   +0.000$ &         --- & km       \\
\hline                                                                                           
$x_{0}$    &   $   +0.107$ & $   +0.097$ & $   +0.107$ & $   +4.359$ & $   +0.100$ & $   +0.000$ & km       \\
$y_{0}$    &   $   -1.145$ & $   -1.049$ & $   -1.145$ & $   -1.050$ & $   -1.057$ & $   +0.000$ & km       \\
$z_{0}$    &   $   +8.503$ & $   +7.781$ & $   +8.503$ & $   +7.777$ & $   +7.793$ & $   +0.000$ & km       \\
$J_{2}$    &   $+0.068726$ & $+0.062803$ & $+0.062864$ & $+0.063890$ & $+0.064671$ & $+0.053887$            \\
$J_{3}$    &   $-0.006287$ & $-0.005294$ & $-0.005751$ & $-0.005386$ & $-0.005396$ & $+0.000000$            \\
$J_{4}$    &   $-0.009600$ & $-0.008835$ & $-0.008781$ & $-0.008847$ & $-0.009927$ & $-0.006243$            \\
$\bar{C}_{20}$   &   $-0.030735$ & $-0.028086$ & $-0.028114$ & $-0.028572$ & $-0.028922$ & $-0.024099$            \\
$\bar{C}_{21}$   &   $+0.000000$ & $+0.000000$ & $+0.000000$ & $+0.000000$ & $+0.000000$ & $+0.000000$            \\
$\bar{S}_{21}$   &   $+0.000000$ & $+0.000000$ & $+0.000000$ & $+0.000000$ & $+0.000000$ & $+0.000000$            \\
$\bar{C}_{22}$   &   $+0.004771$ & $+0.004363$ & $+0.004363$ & $+0.005003$ & $+0.004390$ & $+0.003387$            \\
$\bar{S}_{22}$   &   $+0.000000$ & $+0.000000$ & $+0.000000$ & $+0.000000$ & $+0.000000$ & $+0.000000$            \\
$\bar{C}_{30}$   &   $+0.002376$ & $+0.002001$ & $+0.002174$ & $+0.002036$ & $+0.002040$ & $+0.000000$            \\
$\bar{C}_{31}$   &   $-0.000739$ & $-0.000671$ & $-0.000677$ & $+0.000043$ & $-0.000672$ & $+0.000000$            \\
$\bar{S}_{31}$   &   $+0.000169$ & $+0.000168$ & $+0.000154$ & $-0.000105$ & $+0.000174$ & $+0.000000$            \\
$\bar{C}_{32}$   &   $-0.000926$ & $-0.000828$ & $-0.000848$ & $-0.000871$ & $-0.000829$ & $+0.000000$            \\
$\bar{S}_{32}$   &   $+0.000174$ & $+0.000159$ & $+0.000159$ & $+0.000088$ & $+0.000159$ & $+0.000000$            \\
$\bar{C}_{33}$   &   $+0.000184$ & $+0.000167$ & $+0.000168$ & $-0.000002$ & $+0.000167$ & $+0.000000$            \\
$\bar{S}_{33}$   &   $+0.000521$ & $+0.000474$ & $+0.000476$ & $+0.000452$ & $+0.000488$ & $+0.000000$            \\
$\bar{C}_{40}$   &   $+0.003200$ & $+0.002945$ & $+0.002927$ & $+0.002949$ & $+0.003309$ & $+0.002081$            \\
$\bar{C}_{41}$   &   $+0.000674$ & $+0.000610$ & $+0.000616$ & $+0.000487$ & $+0.000614$ & $+0.000000$            \\
$\bar{S}_{41}$   &   $-0.000142$ & $-0.000131$ & $-0.000130$ & $-0.000053$ & $-0.000112$ & $+0.000000$            \\
$\bar{C}_{42}$   &   $-0.000141$ & $-0.000135$ & $-0.000129$ & $-0.000172$ & $-0.000156$ & $-0.000376$            \\
$\bar{S}_{42}$   &   $+0.000293$ & $+0.000270$ & $+0.000268$ & $+0.000258$ & $+0.000273$ & $+0.000000$            \\
$\bar{C}_{43}$   &   $-0.000521$ & $-0.000475$ & $-0.000476$ & $-0.000433$ & $-0.000475$ & $+0.000000$            \\
$\bar{S}_{43}$   &   $-0.000045$ & $-0.000038$ & $-0.000041$ & $-0.000130$ & $-0.000040$ & $+0.000000$            \\
$\bar{C}_{44}$   &   $+0.000151$ & $+0.000139$ & $+0.000138$ & $+0.000078$ & $+0.000138$ & $+0.000040$            \\
$\bar{S}_{44}$   &   $+0.000261$ & $+0.000238$ & $+0.000238$ & $+0.000230$ & $+0.000253$ & $+0.000000$            \\
\end{tabular}
\end{center}
\caption{Coefficients of each one of the scenarios. 
The top section lists the input parameters:
density of the mantle $\rho_{m}$, density $\rho_{c}$ and radius $R_{c}$ of the core, 
coordinates $\{x_c,y_c,z_c\}$ of the center of the core.
Then follows the resulting
center of mass position $\{x_0,y_0,z_0\}$, the zonal harmonics coefficients $\{J_{l}\}$, 
and the normalized tesseral harmonics coefficients $\{\bar{C}_{lm},\bar{S}_{lm}\}$,
expressed at a reference radius $R_0=300$~km.
Note that for each scenario the coefficients are expressed in a reference system 
with origin in the center of mass
and aligned with the principal axes.
Only the terms up to the 4th degree are listed,
but for most numerical simulations all terms up to the 8th degree were included.
\label{table:scenarios}
}
\end{table*}
}

\end{document}